\documentclass[12pt]{article}
\def\slash#1{\setbox0=\hbox{$#1$}#1\hskip-\wd0\hbox to\wd0{\hss\sl/\/\hss}}
\begin{document}
\baselineskip=20 pt
\def\bea{\begin{eqnarray}}
\def\eea{\end{eqnarray}}
\def\a{\alpha}
\def\b{\beta}
\def\g{\gamma}
\def\m{\mu}
\def\n{\nu}
\def\ep{\epsilon}
\def\e{\eta}
\def\g{\gamma}
\def\p{\partial}
\def\s{\sigma}
\def\r{\rho}
\def\L{\Lambda}
\def\d{\delta}
\def\D{\Delta}
\setcounter{page}{0}
\thispagestyle{empty}
\begin{flushright}
{\bf hep-ph/0211137} \\
November 2002
\end{flushright}

\begin{center}
{\Large{\bf Torsion contraints from the recent precision measurement
of the muon anomaly}}
\end{center}
\vskip 10pT

\begin{center}
{\large\sl Prasanta Das}~\footnote{E-mail:~pdas@mri.ernet.in}
\rm{and} {\large\sl Uma Mahanta}~\footnote{E-mail:~mahanta@mri.ernet.in} \\
\vskip 5pT
{\rm Harish-Chandra Research Institute, \\ 
Chhatnag Road, Jhusi, Allahabad-211019, India .} \\
\vskip 10pT
{\large\sl Sreerup
Raychaudhuri}~\footnote{E-mail:~sreerup@iitk.ac.in} \\
\vskip 5pt
{\rm Department of Physics, Indian Institute of Technology, \\
Kanpur-208016, India.} \\
\end{center}  

\vspace*{0.02in}

\begin{center}
{\bf Abstract}
\end{center}

\noindent
In this paper we consider non-minimal couplings of the Standard Model
fermions to the vector (trace) and axial vector (pseudo-trace) components
of the torsion tensor. We then evaluate the contributions of these vector
and axial vector components to the muon anomaly and use the recent
precision measurement of the muon anomaly to derive constraints on the 
torsion parameters.

\newpage
The Standard Model (SM) has been extremely successful in explaining the
results of all experiments both at low, medium and high energies. In spite
of this phenomenal success a majority of high energy physicists agree that
the SM is at best a renormalizable effective field theory valid in a
restricted energy range since it does not include quantum gravity. Many
theorists further believe that the ultimate and fundamental theory of
Nature will be provided some day by string theory since the low-energy
effective Lagrangian of closed strings reproduces quantum gravity.

The low-energy effective action of closed strings predicts, along with the
a massless graviton, a massless antisymmetric second rank tensor field
$B_{\a\b}$ (Kalb-Ramond tensor) which enters the action\cite{GSW} via its
antisymmetrised derivative $H_{\a\b\g} = \p_{[\a}B_{\b\g]}$. The third 
rank tensor
$H_{\a \b \g}$ is referred to as the torsion field strength. Since torsion
is always predicted by string theory it is highly interesting to
investigate what low-energy observables and phenomenological effects
torsion can produce. Some effects of both heavy (nonpropagating) and
dynamical (propagating) torsion fields have already been considered
\cite{BS} and bounds on the free parameters of the torsion action (torsion
mass and couplings) have been derived using experimental data. In this
paper we shall study the effects of the trace and pseudo-trace components
of the torsion tensor on the anomalous magnetic moment of the muon.

Precision measurements of the muon anomaly $a_\mu = (g_\mu - 2)/2$
currently constitutes one of the most stringent tests for the SM and for
new physics, particularly in the light of the recent and future
promised results from the E821 experiment at BNL\cite{CM}. Recently the E821
collaboration has reported a new improved measurement of $a_\mu$. The
experimental value of the muon anomaly according to this latest
measurement\cite{BEN} is 
$$
a_\mu^{\rm exp} = (11~659~204(7)(4))\times 10^{-10} \ .
$$
The SM prediction, according to the latest and improved
calculation\cite{DWH} of $a_\mu^{\rm hadron}$, is 
$$
a_\mu^{\rm SM} = (11~659~176 \pm 6.7)\times 10^{-10} \ .
$$ 
The experimental value of the muon anomaly
differs, therefore, from the SM prediction by $\delta a_\mu = (28 \pm
10.5)\times 10^{-10}$ {\it i.e.} by about $2.66\sigma$. The BNL
collaboration is expected to reach an eventual precision of 4$\times 10^{-10}$
during its final stage of operation\cite{CM}. If the central
value does not change, this could eventually differ from the SM by about
$7\sigma$ and provide, perhaps, the earliest proof of new physics beyond the
SM.

It is clear, therefore, that it is extremely important to see if ($a$) the
presence of torsion fields can explain why the measured value of the muon
anomaly is different from the SM prediction, and/or ($b$) the highly
accurate measurements from the E821 experiment can be used to constrain
the parameter space of the torsion action. Such an analysis forms the
subject of the present article.

\bigskip

\noindent \S {\sl Torsion couplings to SM fermions}: In dealing with
quantum theory on curved space times with torsion the metric $g_{\m\n}$
and the rank-3 torsion tensor $T^{\a}_{\b\g}$ should be considered as
independent dynamical variables. In this paper, we shall consider the
observable consequences of torsion only and therefore we shall set the
metric to be flat Minkowskian everywhere {\it i.e.} $g_{\m\n} =
\eta_{\m\n}$. 

The torsion field $T^{\a}_{~\b\g}$ is defined\cite{BOS} in terms of the
non-symmetric connection $\widetilde{\Gamma}^{\a}_{~\b\g}$
\bea
T^{\a}_{~\b\g}
= \widetilde{\Gamma}^{\a}_{~\b\g}- \widetilde{\Gamma}^{\a}_{~\g\b} \ .
\eea
For convenience the torsion tensor $T^{\a}_{~\b\g}$ can be divided\cite{BS}
into three irreducible components. These are
\begin{itemize}
\item trace: $T_{\b}=T^{\a}_{~\b\a}$;
\item pseudo-trace: $S^{\n}= \ep^{\a\b\m\n}T_{\a\b\m}$;
\item the third rank tensor $q_{\a\b\g}$, which satisfies the conditions 
$q^{\a}_{~\b\a} = 0$ and $\ep^{\a\b\m\n}q_{\a\b\m}=0$.
\end{itemize}
Clearly $T$ behaves as a vector field and $S$ as an axial vector field.
The simultaneous presence of both $S$ and $T$ as light
dynamical fields could, therefore, be a likely signature of torsion.
However, the simultaneous presence of both $S$ and $T$ as light dynamical
fields in the low-energy effective field theory would lead to serious
problems related to renormalizability. The predictions of the model about
loop-induced corrections like the muon anomaly would then become cut-off
dependent and somewhat uncertain. In this paper we shall, therefore, work
within the simplifying assumption that only one of them appears as a
light, propagating degree of freedom, whereas the other is very heavy and
non-propagating. Hence, we are led to consider the following cases:
\begin{enumerate}
\item $T$ is very heavy and only $S$ appears as the propagating degree of
freedom;
\item $S$ is very heavy and only $T$ appears as the propagating degree of
freedom.
\end{enumerate}
We also assume that the third rank tensor $q_{\a\b\g}$ does not couple
to SM fermions.

We now consider the action for the torsion tensor with non-minimal
couplings to fermions. This assumes the general form
\bea 
S_{\rm torsion} 
& = & \int d^4x 
~\left[ -{1\over 4}S_{\mu\nu} S^{\mu\nu} 
        +{1\over 2} M^2_S S^{\mu}S_{\mu}
        + \eta_S \bar {\psi} \g_{\mu}\g_5 \psi S^{\m} \right] \nonumber \\
& + & \int d^4x 
~\left[ -{1\over 4}T_{\mu\nu} T^{\mu\nu} 
        +{1\over 2} M^2_T T^{\mu}T_{\mu}
        + \eta_T \bar {\psi} \g_{\m}      \psi T^{\m} \right] \ ,
\label{action}
\eea
where $S_{\mu\nu} = \partial_\mu S_\nu - \partial_\nu S_\mu$ and
$T_{\mu\nu} = \partial_\mu T_\nu - \partial_\nu T_\mu$.
Although in the above Lagrangian we have included the kinetic energies for
both $S$ and $T$, it should be kept in mind that for the underlying theory
to be renormalizable, only one of them should considered as propagating.
For phenomenological purposes only the first or the second line of
Eqn.(\ref{action}) is relevant, never both. 

We note that in the minimal coupling scheme $\eta_T = 0$. A non-zero
$\eta_T$, therefore, represents purely non-minimal effects. Moreover,
since the vector current is exactly conserved (CVC), the mass $M_T$ of
$T_{\m}$ can be either zero or non-zero without affecting the
renormalizability of the model. On the other hand, partial conservation of
the axial vector current (PCAC) implies that $M_S$ must be non-zero for
the above action to be renormalizable.

\bigskip

\noindent \S {\sl Torsion contributions to the muon anomaly}: In the
following we shall estimate the muon anomaly contribution due to the
torsion tensor separately for the two cases mentioned above.

\noindent ($a$) In this case, only the axial vector field ($S_{\mu}$)
appears as the propagating degree of freedom and the muon anomaly
contribution due to torsion arises from the one-loop diagram in
Fig.~1($a$). Using the Gordon identity to express the vertex correction as
a sum of a vector current and spin current, we finally get
\bea
a_\mu^{(S)} = {m_\mu^2\eta_S^2\over 4\pi^2}
\int_0^1 dx ~\frac{x(1-x)(x-4) - 2x^3{m_\mu^2\over M_S^2}}
                  {m_\mu^2x^2 + M_S^2(1-x)} \ .
\label{Scontrib}
\eea
Two particular limits of $M_S$ are particularly interesting, namely $M_S
\gg m_\mu$ and $M_S \approx m_\mu$. For $M_S \gg m_\mu$, the above results
simplifies to
\bea
a_\mu^{(S)} \simeq -\frac{5}{3} \left(\frac{m_\mu \eta_S}{2\pi M_S}\right)^2
\ ,
\eea
while, for $M_S \simeq m_\mu$ we get
\bea
a_\mu^{(S)} \simeq 
- \frac{\eta_S^2}{4\pi^2}\int_0^1 dx ~\frac{4x-5x^2+3x^3}{1-x+x^2}
= - 1.3138 \left(\frac{\eta_S}{2\pi}\right)^2 \ .
\eea
Note that the contribution of $S$ to the muon anomaly is always negative.

\begin{figure}[htb]
\begin{center}
\vspace*{3.0in}
  \relax\noindent\hskip -5.4in\relax{\includegraphics{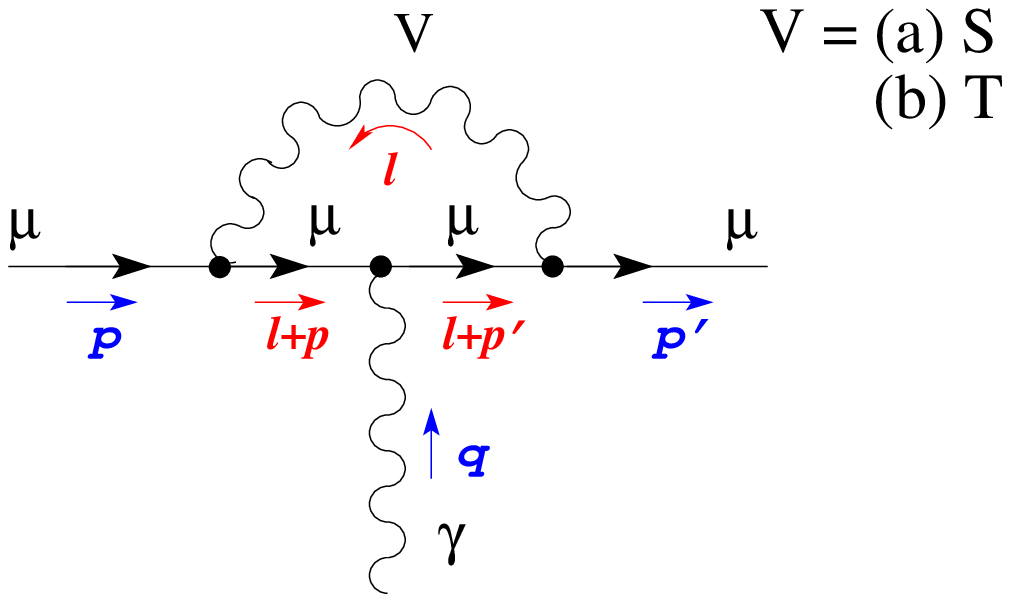}}
\end{center}
\end{figure}
\vspace*{-1.0in}
\noindent {\bf Figure 1.} {\footnotesize 
One-loop Feynman diagrams that contribute to the muon anomaly through 
exchange of \\ \hspace*{0.8in} ($a$) $S$ and ($b$) $T$ fields.}
\vskip 10pt

\noindent 
($b$) In this case only the trace ($T^{\m}$) of the torsion tensor appears
as the propagating dynamical field and the muon anomaly contribution
due to torsion arises from Fig. 1($b$). Following the usual procedure
we then find
\bea
a_\mu^{(T)} 
= {m_\mu^2\eta_T^2\over 4\pi^2}\int_0^1 dx 
\frac{x(1-x)}{x^2m_\mu^2 +(1-x)M_T^2} \ .
\label{Tcontrib}
\eea
For $M_T \gg m_\mu$ we get
\bea
a_\mu^{(T)} \simeq  \frac{1}{3} \left(\frac{m_\mu \eta_T}{2\pi M_T}\right)^2
\eea
while for $M_T \simeq m_\mu$,
\bea
a_\mu^{(T)} \simeq
{\eta_T^2\over 4\pi^2}\int_0^1 dx ~\frac{x(1-x)}{1-x+x^2}
= 0.2092 \left(\frac{\eta_T}{2\pi}\right)^2 \ .
\eea
The muon anomaly contribution due to $T$ is, therefore, always positive.

\bigskip

\noindent \S {\sl Determination of constraints}: The above formulae can be
now used in conjuction with the experimental results to obtain constraints
on the parameter space of torsion fields.  We first consider the
constraints that arise when the torsion mass is much greater than the muon
mass. Under this condition the muon anomaly due to $S$ can be expressed as
\bea
a_\mu^{(S)} = -{5\over 3\pi} \left({m_\mu\over \L_S}\right)^2
\eea
where $\L_S$ is an effective scale given by $\L_S^2=4\pi M_S^2/\eta_S^2$. 
Since the experimental result already shows a positive deviation
($2.66\sigma$) from the SM value, the effects of $S$ would cause an even
further deviation. We can parameterize this excess deviation by
\bea
\xi_S={\vert\d a_\mu^{\rm exp}({\rm C.V.})- a_\mu^{(S)} \vert
\over \D \left(\d a_\mu^{\rm exp}\right)}
\eea
where 
$$
\d a_\mu^{\rm exp} 
= \d a_\mu^{\rm exp}({\rm C.V.}) \pm \D (\d a_\mu^{\rm exp}) \ ,
$$ 
C.V. standing for the experimental central value. In the decoupling limit,
$\L_S\rightarrow \infty$, $a_\mu^{(S)} \rightarrow 0$ and
$\xi_S\rightarrow 2.66$ which is the SM result. The presence of $S$
 cannot, therefore, account for the present experimental value of the muon
anomaly for $\xi_S <2.66$. The variation of $\xi_S$ with $\L_S$ is shown
in Fig. 2($a$), together with the $3\sigma$ bound which corresponds to 
\bea
M_S > \eta_S ~(1.13 ~{\rm TeV})
\eea
A similar analysis can be done for the vector component $T$. For
$M_T \gg m_\mu$ the muon anomaly contribution due to $T$ can be written as
\bea
a_\mu^{(T)} ={1\over 3\pi}\left({m_\mu\over \L_T}\right)^2
\eea
where $\L_T$ is another effective scale given by $\L_T^2 = 
4\pi M_T^2/\eta_T^2$. Since the contribution of $T$ to the muon anomaly is
positive, its presence {\it can constitute an explanation} for the muon
anomaly both for positive and negative deviations from 
$\d a_\mu^{\rm exp}({\rm C.V.})$. As before, we can parametrize the
contribution to the muon anomaly by
\bea
\xi_T = {\vert\d a_\mu^{\rm exp}({\rm C.V.})- a_\mu^{(T)} \vert
\over \D \left(\d a_\mu^{\rm exp}\right)}
\eea
where $\xi_T$ denotes the number of standard deviations by which
$a_\mu^{(T)}$ differs from $\d a_\mu^{\rm exp}({\rm C.V.})$. The variation
of $\xi_T$ with $\L_T$ is shown in Fig. 2($b$) along with the upper and
lower bounds at 1$\sigma$ and 2$\sigma$ and the lower bound at 3$\sigma$.
The left (right) branch corresponds to positive (negative) deviations from
$\d a_\mu^{\rm exp}({\rm C.V.})$.  For $\L_T \approx $ 792 GeV, $\xi_T$
vanishes --- indicating this is the most favored value if torsion is the
cause for the present muon anomaly. In fact, we obtain the following
bounds
\bea
\L_T &=& 792^{+209}_{-117}~~~{\rm GeV~~~at}~~~ 1\sigma \nonumber \\
     &=& 792^{+791}_{-194}~~~{\rm GeV~~~at}~~~ 2\sigma \nonumber \\
     &>& 543         ~~~~~~~~{\rm GeV~~~at}~~~ 3\sigma
\eea  

\begin{figure}[htb]
\begin{center}
\vspace*{3.7in}
  \relax\noindent\hskip -6.4in\relax{\includegraphics{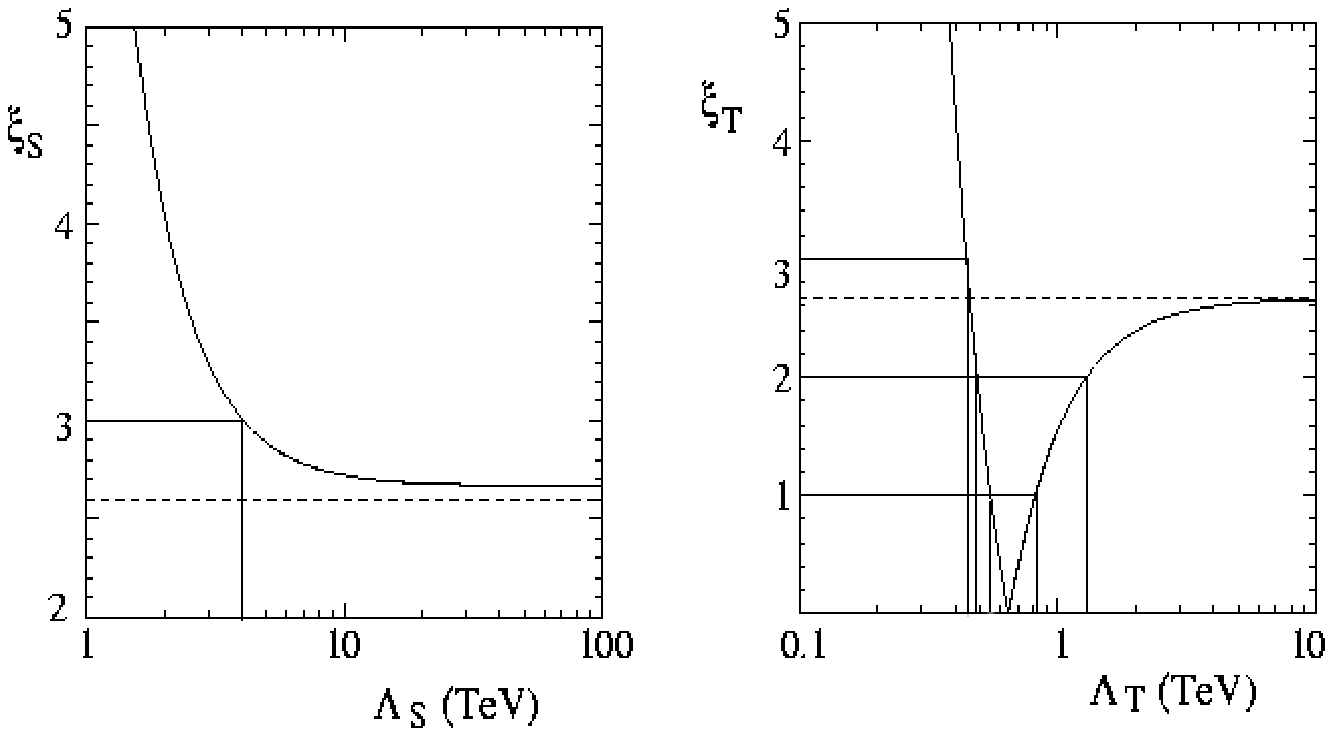}}
\end{center}
\end{figure}
\vspace*{-1.0in}
\noindent {\bf Figure 2.} {\footnotesize
Illustrating constraints on the torsion parameter space from the muon anomaly
with \\ \hspace*{0.8in} exchange of ($a$) $S$ and ($b$) $T$ fields in the 
limit $M_{S,T} \gg m_\mu$.}
\vskip 10pt

We now examine the situation when $M_{S, T} \approx m_\mu$. In this case
the muon anomaly due to torsion cannot be parameterised in terms
of a single parameter $\L_{S, T}$ as we have done in the above discussion.
Here the integrals in Eqns. (\ref{Scontrib}) and (\ref{Tcontrib}) have to
be evaluated exactly. Noting that $a_\mu^{(S)}$ is always negative and
only a $3\sigma$ bound can be obtained, while the positive $a_\mu^{(T)}$
allows for both upper and lower bounds at 1$\sigma$ and 2$\sigma$, we
require
\bea
\vert a_\mu^{(S)}\vert 
& \le  & 3\D (\d a_\mu^{\rm exp})- \d a_\mu^{\rm exp}({\rm C.V.}) 
= 3.5\times 10^{-10}~~~~~{\rm at}~3\sigma \nonumber\\
\vert a_\mu^{(T)}-\d a_\mu^{\rm exp}({\rm C.V.})\vert 
& \le & {\rm n} \D(\d a_\mu^{\rm exp}) \hspace*{1.0in}
=10.5n\times 10^{-10}~~{\rm at}~{\rm n}\sigma 
\eea
Using these conditions we can now constrain the $\eta_S$--$M_S$ and
$\eta_T$--$M_T$ planes. These are shown in Figs. 3($a$) and ($b$)
respectively. Broad hatching represents the allowed region at $3\sigma$,
close hatching represents the allowed region at $2\sigma$ and cross
hatching represents the allowed region at $1\sigma$. We have considered
the range $M_{S, T} = 50$~MeV -- 1~GeV.  Beyond a GeV it is reasonable to
use the approximation $M_{S,T} \gg m_\mu$, which has already been
discussed. We note that for light $S$ or $T$ the muon anomaly result
requires $\eta_S$ and $\eta_T$ to be very small. This constitutes an
explanation as to why torsion effects have not been seen so far in
low-energy precision experiments.

\begin{figure}[htb]
\begin{center}
\vspace*{3.7in}
  \relax\noindent\hskip -6.4in\relax{\includegraphics{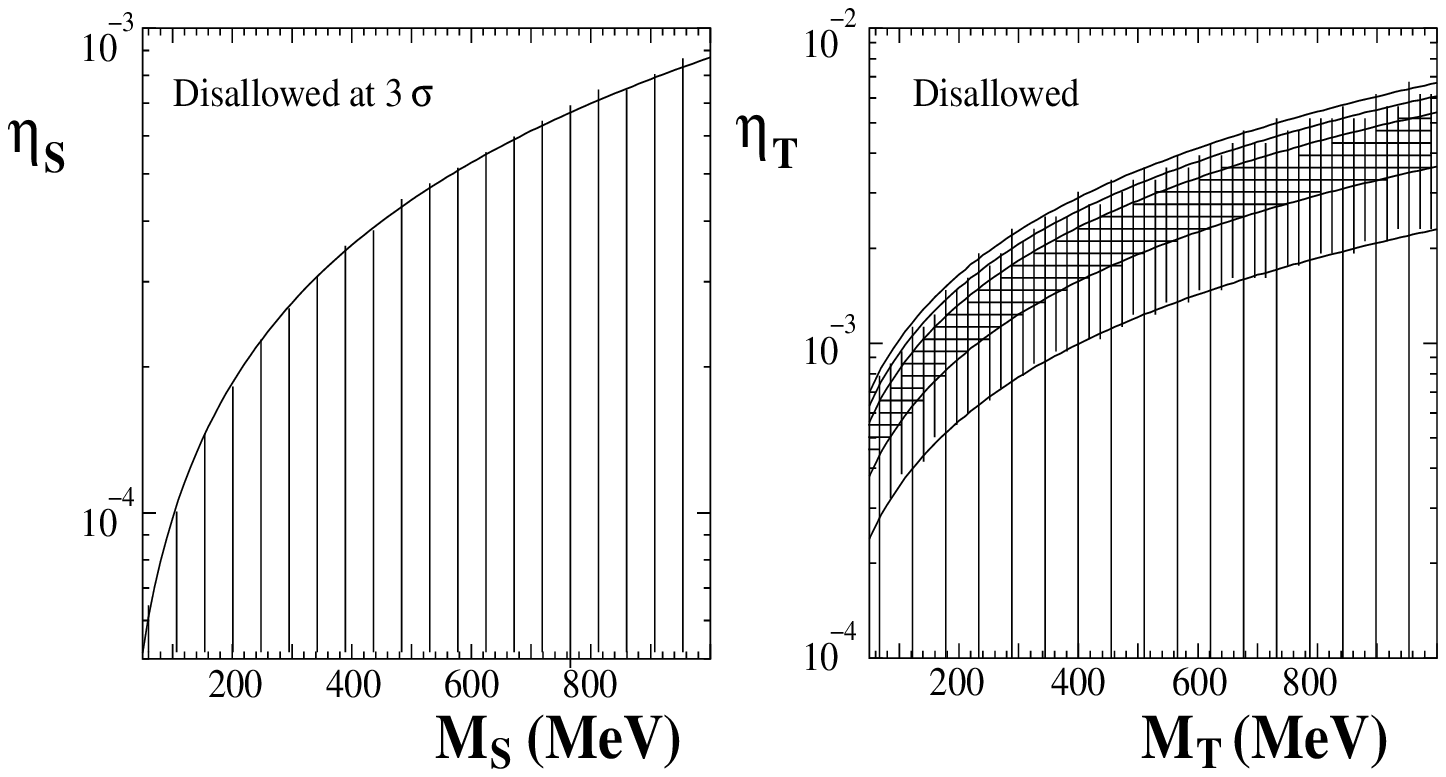}}
\end{center}
\end{figure}
\vspace*{-0.6in}
\noindent {\bf Figure 3.} {\footnotesize
Illustrating constraints on the torsion parameter space from the muon anomaly
with \\ \hspace*{0.8in} exchange of ($a$) $S$ and ($b$) $T$ fields in the
limit $M_{S,T} \approx m_\mu$.}
\vskip 10pt

\bigskip

\noindent \S {\sl Discussion of results}: It is interesting to compare the
bounds on torsion-fermion couplings and torsion masses obtained from the
muon anomaly with the existing direct bounds obtained from collider data.
Since collider bounds are available only for the $S$ field and that too
for $M_S \gg m_\mu$, we must perforce confine the discussion to this case
only.  Using LEP-1.5 data on $A^{(e)}_{FB}$, Belyaev and Shapiro\cite{BS}
have found a lower bound $M_S > 2.1$~ TeV for $\eta_S \simeq 1$.  On the
other hand, from the muon anomaly we obtain a lower bound $M_S \simeq
1.13$~TeV for $\eta_S \simeq 1$. The constraints obtained from the muon
anomaly are therefore slightly weaker --- but nevertheless comparable ---
with the direct collider bounds. The bounds obtained from the muon anomaly
would however improve once the BNL collaboration enters the final stage of
operation. It would also be interesting to update\cite{us} the collider
bounds on torsion parameters in the light of more recent data from LEP-2
and from Run-I of the Tevatron.

\begin{figure}[htb]
\begin{center}
\vspace*{4.0in}
  \relax\noindent\hskip -6.4in\relax{\includegraphics{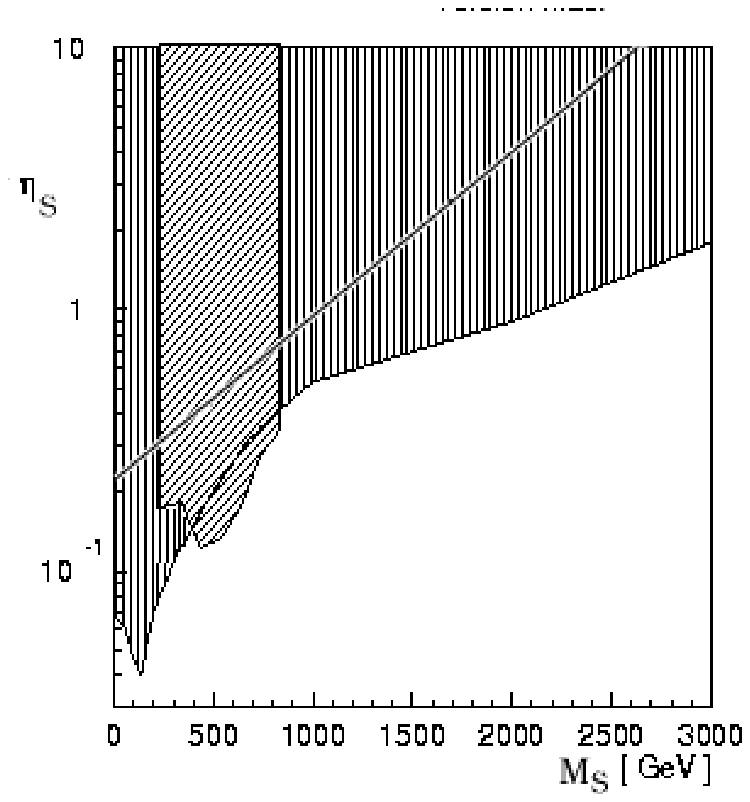}}
\end{center}
\end{figure}
\vspace*{-1.2in}
\noindent {\bf Figure 4.} {\footnotesize
Comparing direct constraints on the parameter space for the $S$ field with
those obtained from the muon anomaly. Vertical hatching represents the
region disallowed by $A_{FB}^{(e)}$ and oblique hatching represent the CDF
constraint, both as obtained by Belyaev and Shapiro\cite{BS}. The straight
line corresponds to the $3\sigma$ bound from the muon anomaly.} 
\vskip 10pt

{\footnotesize\noindent \S {\sl Acknowledgments:}
UM would like to thank Pankaj Jain and the Department of Physics, I.I.T.
Kanpur, for hospitality while a part of this work was being done. SR
wishes to thank the Harish-Chandra Research Institute for hospitality
while the paper was being written.}


\end{document}